\documentstyle[eqsecnum,aps,epsfig]{revtex}

\begin{document}
\draft
\preprint{HEP/123-qed}
\title{In-medium properties of the D$_{13}(1520)$ nucleon resonance}
\author{
B.~Krusche$^1$, 
J.~Ahrens$^2$, 
R.~Beck$^2$,
I.J.D.~MacGregor$^3$,
J.C.~McGeorge$^3$,
V.~Metag$^4$,
H.~Str\"oher$^5$}
\address{
$1$ Department of Physics and Astronomy, University of Basel,
Ch-4056 Basel, Switzerland\\
$^2$Institut f\"ur Kernphysik, 
Johannes-Gutenberg-Universit\"at
Mainz, D-55099 Mainz, Germany\\
$^3$Department of Physics and Astronomy, University of Glasgow,
Glasgow G128QQ, UK\\
$^4$II. Physikalisches Institut, Universit\"at Giessen, 
D-35392 Giessen, Germany\\
$^5$Institut f\"ur Kernphysik,
Forschungszentrum J\"ulich GmbH,
52425 J\"ulich, Germany
}
\date{\today}
\maketitle
\begin{abstract}
The in-medium properties of the D$_{13}(1520)$ nucleon resonance were studied
via photoproduction of $\pi^o$-mesons from nuclei (C, Ca,
Nb, Pb) with the TAPS-detector at the Mainz MAMI accelerator.
The inclusive (single and multiple pion production) data disagree
with model predictions which explain the disappearance of the second
resonance bump in total photoabsorption via a medium modification of the
D$_{13}\rightarrow \mbox{N}\rho$ decay.  
The exclusive single $\pi^o$ production data show no broadening of the 
resonance structure beyond Fermi smearing. Although due to final state
interaction effects pion production tests the nucleus at densities less than
the average nuclear density, both results together cast doubt on attempts 
to  explain the vanishing of the second resonance bump for heavy nuclei
by a broadening of the D$_{13}$-resonance in nuclear matter. 
\end{abstract}
\pacs{PACS numbers: 
13.60.Le, 14.20.Gk, 14.40.Aq, 25.20.Lj
}

\narrowtext

The low lying $\mbox{N}^{\star}$ resonances, excited states of the nucleon with
isospin $I=1/2$, comprise the states P$_{11}(1440)$, D$_{13}(1520)$ and 
S$_{11}(1535)$ in the well known nomenclature \cite{PDG}. They are for example 
excited by photons with energies between 600 and 900 MeV. Since their decay 
widths are large compared to their spacing they overlap and form one single 
enhancement usually called the second resonance region. Due to their different 
couplings to the initial photon - nucleon and the final meson - nucleon states 
they can nevertheless be separated to a large extent:
The production of $\eta$-mesons
proceeds almost exclusively via the excitation of the S$_{11}$ resonance,
while the largest resonance contributions to single and double pion 
production come from the D$_{13}$ resonance. Using this selectivity,
the properties of the resonances, when excited on the free proton
or quasifree neutron, have been studied in much detail during the last 
few years via $\eta$-photoproduction 
\cite{Krusche_1,Krusche_2,Bock,Ajaka,Krusche_3,Hoffmann,Hejny} 
and single and double pion photoproduction reactions 
\cite{Braghieri,Tejedor,Haerter,Zabrodin,Krusche_4,Kleber,Wolf}.
The excellent quality of the recent data sets 
allowed precise determinations of the resonance properties, e.g. the extraction
of a 0.05\% - 0.08\% branching ratio for the 
D$_{13}\rightarrow \mbox{N}\eta$ decay \cite{Tiator}.

However, much less is yet known about the behavior of the isobars 
inside the nuclear medium, where a number of modifications may arise. 
The most trivial is the broadening of the excitation functions due to 
Fermi motion. The decay of the resonances is modified by Pauli-blocking
of final states, which reduces the resonance widths, and by additional decay
channels like $\mbox{N}^{\star}\mbox{N}\rightarrow \mbox{NN}$ which cause 
the so-called collisional broadening. Both effects cancel to some extent and 
it is a priori not clear which one will dominate.
A very exciting possibility is that the resonance
widths could be sensitive to in-medium mass modifications of mesons arising 
from chiral restoration effects \cite{Brown,Mosel}. The D$_{13}$-resonance for 
example has a 15 - 25\% decay branch to the $\mbox{N}\rho$-channel \cite{PDG}, 
which is only fed from the low energy tail of the $\rho$ mass distribution.

\newpage
\vspace*{8.cm}
 
\noindent{This} means that a broadening or a downward shift of the 
$\rho$-mass distribution
inside the nuclear medium could have significant effects on the D$_{13}$ width.

The first experimental investigation of the second resonance region for nuclei
was done with total photoabsorption. The surprising results showed
an almost complete absence of the resonance bump for $^{4}$He and 
heavier nuclei \cite{Frommhold,Bianchi,MacCormick},
which up to now has not been understood. 
\begin{figure}[h]
\begin{center}
\epsfig{figure=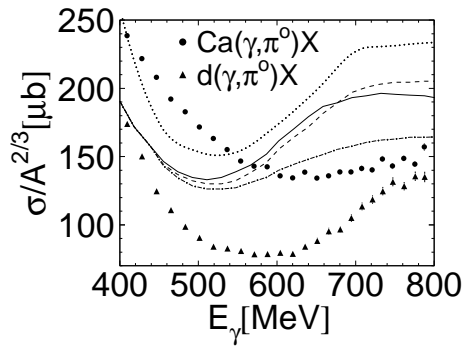,width=6cm}\\[0.5cm]
\caption{Total inclusive $\pi^o$ photoproduction cross section from $^2$H 
\protect\cite{Krusche_4} and 
Ca scaled by $A^{2/3}$ compared to model calculations \protect\cite{Lehr}
for $\mbox{Ca}(\gamma ,\pi^o)X$.
Dotted curve: BUU, dashed curve: BUU with $\Delta$ collision width from $\Delta$-hole model,
solid curve: like dashed curve but modified D$_{13}(1520)\rightarrow$N$\rho$
width, dash-dotted: like dashed curve but additional 300 MeV collision width
of D$_{13}$.
\label{fig:1}
}
\end{center}
\end{figure}
The problem is illustrated in
Fig. \ref{fig:1} where the total inclusive $\pi^o$-production cross section 
from Ca from the present experiment is compared to the reaction on the 
deuteron and to model predictions. The deuteron data show a clear bump around
800 MeV, but the calcium data are rather flat. However, the 
predictions for Ca from the BUU model \cite{Lehr},
which takes into account Fermi smearing, Pauli-blocking and collisional 
broadening still exhibit the bump structure.  Not even the modification of the 
D$_{13}$ decay by medium effects of the $\rho$-meson (solid curve) 
improves the situation. Only an arbitrary and probably unrealistic broadening 
\cite{Lehr} of the 
D$_{13}$-resonance by 300 MeV produces a significant suppression.
Note that the underestimation of the cross section at low incident
photon energies is attributed in \cite{Lehr} to many-body absorption processes
of the photon which are not included in the model.

Total photoabsorption has the advantage, that no final state
interaction effects (FSI) must be accounted for so that the entire nuclear
volume is tested. However, many different reaction channels do contribute to
this reaction so that it is impossible to test the behavior of individual
resonances. Even worse, some of the reaction channels responsible for 
the structure are not strongly related
to the excitation of states from the second resonance region. 

Part of the problem is that most of the rise of the cross section towards 
the resonance bump stems from the opening of the double pion production channels
\cite{Krusche_5}. The largest contribution is due to the $\pi^+\pi^-$ final 
state. It is well known \cite{Tejedor} that this reaction is dominated by 
$\Delta$-Kroll-Rudermann and $\Delta$-pion-pole terms, i.e. it mainly involves 
the excitation of the P$_{33}(1232)$ $\Delta$-resonance rather than the 
states from the second resonance region.
 
It is therefore desirable to study this region with exclusive reactions which
allow the investigation of individual resonances, even at the expense that FSI
effects complicate the interpretation. In a recent study \cite{Roebig} we 
used $\eta$-photoproduction for an investigation of the in-medium properties of
the S$_{11}(1535)$. In this case we did not find any unexplained depletion of
the in-medium resonance strength. The data were in excellent agreement with 
predictions from the BUU-model \cite{Effenberger} and other models
(e.g. \cite{Carrasco}), taking into account Fermi smearing, Pauli-blocking, 
collisional broadening and FSI effects. Since the energy range extended just up
to the resonance maximum, it was not possible to deduce the in-medium width of
the resonance. In the meantime Yorita et al. \cite{Yorita} have studied
quasifree $\eta$-photoproduction from carbon over a larger energy range 
and also found no significant broadening of the S$_{11}$ resonance. Again the 
data are in fairly good agreement with model expectations.

The above results do not preclude broadening as the explanation for 
the absence of the second resonance bump because the total contribution of the 
S$_{11}$-resonance to the bump structure is quite small.
Furthermore due to the location of the $\eta$-production threshold 
in the low energy tail of the resonance, the effects of nuclear Fermi motion 
have a drastic influence on the excitation curve so that any extraction of
the in-medium width requires a lot of modelling. In the present
work we have therefore investigated the dominant D$_{13}(1520)$ resonance in the nuclear
medium via quasifree single $\pi^o$-photoproduction. 

The experiments using C-, Ca-, Nb- and Pb-targets
were carried out at the Glasgow tagged photon facility installed at the Mainz
Microton (MAMI) with the TAPS-detector. Details of the experimental setup and
the data analysis are summarised in \cite{Krusche_4}. The neutral pions were
identified via an invariant mass analysis and quasifree single
$\pi^o$-production was selected by a missing energy analysis as in
\cite{Krusche_4}. The stronger broadening of the structures in the missing 
energy spectra for nuclei heavier than deuterium  
was compensated by more restrictive cuts, so that contamination from
multiple meson production was excluded.  
 
\begin{figure}[h]
\begin{center}
\epsfig{figure=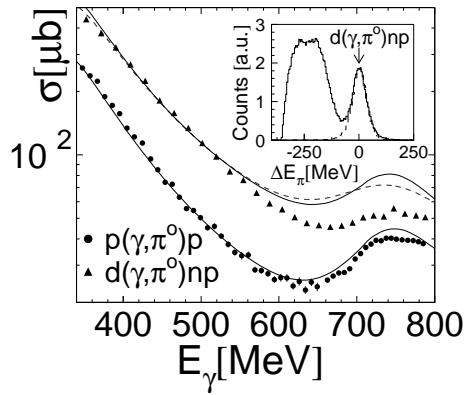,width=6cm}\\[0.5cm]
\caption{Comparison of the measured total cross sections for $p(\gamma,\pi^o)p$
and $d(\gamma,\pi^o)np$ to the MAID analysis \protect\cite{MAID}.
Proton case, solid line: MAID analysis.
Deuteron case, solid curve: sum of proton and neutron MAID cross section; 
dashed curve: folded with momentum distribution of bound nucleons.
Insert: missing energy spectrum of the reaction $d(\gamma ,\pi^o)X$ for incident
photon energies 600 - 800 MeV used to separate single $\pi^o$-production from
multiple pion production reactions. Dashed line: Monte Carlo simulation
for single $\pi^o$-production (see \protect\cite{Krusche_4}).  
\label{fig:3}
}
\end{center}
\end{figure}
The total single $\pi^o$ production cross section for the proton and the 
deuteron \cite{Krusche_4} are shown in Fig. \ref{fig:3}.
The insert shows for the deuteron the separation of quasifree 
single $\pi^o$-production in missing energy from multiple pion production 
processes ($\pi^o\pi^o$,
$\pi^{\pm}\pi^o$, $\eta\rightarrow 3\pi^o$, $\eta\rightarrow\pi^o\pi^+\pi^-$),
which contribute to the inclusive cross section.
In the main part of the figure the proton and deuteron cross
sections are compared to the expectation from a unitary isobar analysis
of pion photoproduction (MAID) \cite{MAID}. The data for the proton 
are very well reproduced. For the deuteron, we have taken the sum of the 
proton and neutron cross sections from MAID (full curve in Fig. \ref{fig:3})
and folded this cross section with the momentum distribution of the nucleons
bound in the deuteron (dashed curve). The momentum distribution was derived 
from a parametrization of the deuteron wave function \cite{Lacombe}.
The prediction for the deuteron cross section agrees very well
with the data in the tail of the $\Delta$-resonance, but it significantly
overestimates the cross section in the D$_{13}$ region. This result is very
surprising since we are dealing with {\it quasifree} pion production, for  
which the large momentum mismatch between participant and spectator nucleon
is expected to suppress any interference terms between the two nucleons.
At present it is not clear if this problem is related to the input
used for the elementary cross section of the $n(\gamma ,\pi^o)n$ reaction or if
the incoherent addition of the Fermi smeared
elementary cross sections, which in the same energy region works excellently
e.g. for $\eta$-photoproduction \cite{Krusche_3,Hejny}, is not a good 
approximation.   
Precise measurements of $\pi^o$-photoproduction from the deuteron with 
coincident detection of the recoil nucleons and more refined model 
calculations are necessary to solve this problem.
However, no matter what the nature of the problem is, we like to stress
that this finding has important consequences for the extraction of resonance
in-medium properties from a comparison to model predictions.
Models like the BUU \cite{Lehr} must rely
on the assumption that the total cross section from nuclei
before taking into account in-medium and FSI effects is the incoherent sum of
the known proton and neutron cross sections, which in this case is not even true
for the deuteron.

\begin{figure}[h]
\begin{center}
\epsfig{figure=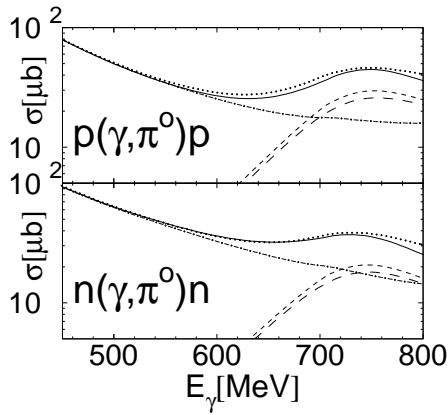,width=5.8cm}\\[0.5cm]
\caption{Decomposition of single $\pi^o$ photoproduction in the second resonance
region from MAID2000 
\protect\cite{MAID}. Full lines: cross section $\sigma_{\pi^o}$ for full 
model; dash-dotted curves: cross section $\sigma_{nr}$ without contribution
from D$_{13}$ and S$_{11}$ resonances, short-dashed curves: cross section
$\sigma_{r}$ for excitation of D$_{13}$ and S$_{11}$ only, long-dashed curves:
D$_{13}$ only, dotted curves: $\sigma_{nr}+\sigma_{r}$. 
\label{fig:4}
}
\end{center}
\end{figure}

\begin{figure}[h]
\begin{center}
\epsfig{figure=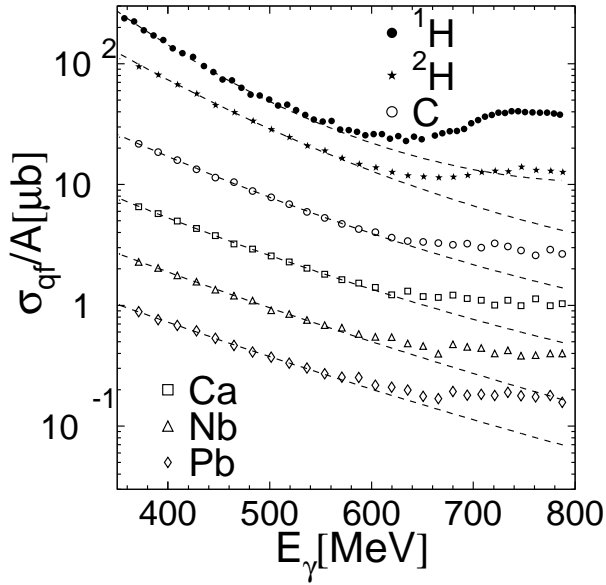,width=7.9cm}\\[0.5cm]
\caption{Total cross section per nucleon for single $\pi^o$ 
photoproduction in the second resonance region for the nucleon and for nuclei.
The scale corresponds to the proton data, the other data are scaled down
by factors 2,4,8,16,32. The dashed curves are fits to the data in the energy
range 350 - 550 MeV.
\label{fig:5} 
}
\end{center}
\end{figure}

For a more quantitative analysis of the D$_{13}$-excitation in nuclei we have
decomposed the cross sections into a resonance and a background part. In
principle such a decomposition requires a multipole analysis wich takes into
account resonance - background interference terms. However, interference terms 
are small in this case as demonstrated in Fig. \ref{fig:4}, which shows
the proton and neutron cross sections calculated with MAID \cite{MAID}.
The single $\pi^{o}$ production cross sections
taking into account all resonances and background terms ($\sigma_{\pi^o}$)
are very similar to the sum of the separate cross sections $\sigma_r$ 
(excitation of the S$_{11}$ and D$_{13}$ resonances only) and $\sigma_{nr}$
(everything except S$_{11}$ and D$_{13}$ excitation).
In the following we do not attempt to separate the
contribution from the two resonances but one should keep in mind that the
resonance part is dominated by the excitation of the D$_{13}$
(see Fig. \ref{fig:4}). The decomposition of the measured cross sections is
shown in Fig. \ref{fig:5}. 
The background part coming from the tail of the
$\Delta$-resonance, the contribution of the P$_{11}$-resonance, nucleon Born
terms and vector meson exchange was fitted with a function of the type:
\begin{equation}
\sigma\propto e^{(aE_{\gamma}^2+bE_{\gamma})} 
\eqnum{1}
\end{equation}
with $a$ and $b$ as free parameters. 
It is obvious from the figure that the resonance contribution for the data
for heavier nuclei is not qualitatively different from the deuteron case.

\begin{figure}[h] 
\begin{center}
\epsfig{figure=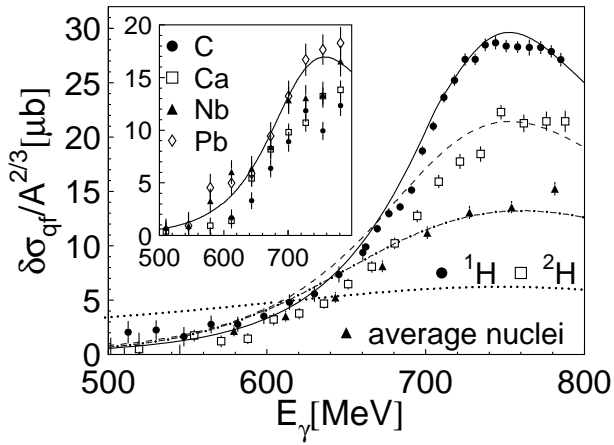,width=8cm}\\[0.5cm]
\caption{Main figure: differences between measured cross sections and 
fits shown in Fig. \protect\ref{fig:5}
scaled by $A^{2/3}$. Full curve: MAID prediction for excitation of the
D$_{13}$ and S$_{11}$ resonances on the proton; dashed curve: MAID
proton - neutron average folded with momentum distribution for deuteron;
dash-dotted: curve folded with momentum distribution for medium weight nuclei
(both scaled to the data).
Dotted curve: Breit-Wigner curve for the D$_{13}$ resonance with 300 MeV width.
Insert: individual nuclear data and prediction from deuteron cross section
(solid curve, see text).
\label{fig:6}
}
\end{center}
\end{figure}

The differences between measured cross sections and fits are
shown in Fig. \ref{fig:6}.
In the main figure the resonance contributions for the proton, the
deuteron and the average for the heavier nuclei are compared to the MAID
predictions for the D$_{13}$ and S$_{11}$ contributions. For the deuteron 
and the heavier nuclei the MAID average of proton and neutron cross sections 
folded with the proper momentum distributions is scaled to the data.
Obviously no broadening of the
resonance structure beyond Fermi smearing is observed. A D$_{13}$ resonance
broadened to 300 MeV as used in the BUU calculations \cite{Lehr} for the 
inclusive data (see Fig. \ref{fig:1}) is clearly ruled out, the data correspond
rather to BW-curves with a width around 100 MeV.

Finally, we have investigated if the strength of the resonance signal 
for the nuclei is consistent with the deuteron case. For this purpose we have
folded the MAID proton cross section for resonance excitation with the 
deuteron Fermi motion and compared the result to the measured deuteron cross 
section. Agreement is obtained for
$\sigma_n(\mbox{D}_{13})/\sigma_p(\mbox{D}_{13})\approx 1/3$, while the ratio 
obtained from
a direct comparison of MAID proton and neutron cross sections is 2/3.
We have then adopted the 1/3 ratio, folded $(1+1/3)\sigma_p/2$ with a
typical nuclear momentum distribution and compared the result to the
nuclear data scaled to $A^{2/3}$, which in the simplest approximation accounts 
for the FSI effects (see Fig. \ref{fig:6}, insert). The agreement of this 
approximation with the data is quite good.

At this point we must clarify a crucial aspect of our results. The approximate 
scaling of the cross sections with $A^{2/3}$ indicates FSI effects. This
means that in contrast to total photoabsorption not the entire nuclear volume 
is probed. BUU-model calculations \cite{Hombach} indicate 
that e.g. for $^{208}$Pb in the $\Delta$-resonance region {\it observed}
pions are predominantly produced in a surface region 
where the nuclear density drops from 0.8$\rho_o$ to 0.4$\rho_o$ 
($\rho_{o}$: normal nuclear density). 
However, suppression of the resonance bump in 
total photoabsorption reactions occurs already for $^4$He \cite{MacCormick} and
does not change from very light nuclei like lithium and beryllium up to very 
heavy ones like uranium. This excludes a strong density dependence of the 
effect.
Furthermore, it is clear that the models without a strong broadening 
of the D$_{13}$ resonance overestimate our inclusive pion data 
(see Fig. \ref{fig:1}). 
However, these data are subject to FSI in the same way as the exclusive data. 
This can be shown by fitting the mass dependence of the cross section from 
carbon to lead with a simple $\propto A^{\alpha}$ law. The
results in the energy range 720 - 790 MeV are: $\alpha=0.791\pm 0.005$
(inclusive pion data), $\alpha=0.74\pm 0.01$ 
(exclusive pion data, Fig. \ref{fig:5}), and $\alpha=0.81\pm 0.05$ 
(D$_{13}$ part only, Fig. \ref{fig:6}, insert).
This means that the inclusive and exclusive data probe the nuclei at comparable 
densities and consequently a broadening of the D$_{13}$ resonance is 
ruled out as an explanation for the overestimation. 
It is thus evident that the models miss some other effect
which must be understood before the results from total photoabsorption can be
used as evidence for an in-medium resonance broadening. 

In summary, investigating quasifree $\pi^o$ photoproduction we have
found a strong quenching but no broadening of the D$_{13}$-resonance structure
for the deuteron with respect to the proton. However, for heavy nuclei we found
no indication of a broadening or a suppression of the D$_{13}$ structure with 
respect to
the deuteron. Since so far model predictions agree with the pion photoproduction
data only under the assumption of a strong broadening of the resonance, 
other effects seem to be missing in the models. This casts also doubt 
on the interpretation of the total photoabsorption data via resonance 
broadening in the framework of the same models.

\acknowledgments

We wish to acknowledge the outstanding support of the accelerator group of
MAMI, as well as many other scientists and technicians of the Institute
f\"ur Kernphysik at the University of Mainz. 
This work was supported by 
Deutsche Forschungsgemeinschaft (SFB 201)
and the UK Science and Engineering Research Council.


\begin{references}
\bibitem{Krusche_1} B. Krusche {\it et al.}, Phys. Rev. Lett. {\bf 74} 3736
(1995).
\bibitem{Krusche_2} B. Krusche {\it et al.}, Phys. Lett. {\bf B397} 171 (1997).
\bibitem{Bock} A. Bock {\it et al.}, Phys. Rev. Lett. {\bf 81} 534 (1998).
\bibitem{Ajaka} J. Ajaka {\it et al.}, Phys. Rev. Lett. {\bf 81} 1797 (1998).
\bibitem{Krusche_3} B. Krusche {\it et al.}, Phys. Lett. {\bf B358} 40 (1995).
\bibitem{Hoffmann} P. Hoffmann-Rothe {\it et al.}, Phys. Rev. Lett. {\bf 78} 4697
(1997).
\bibitem{Hejny} V. Hejny {\it et al.}, Eur. Phys. J. {\bf A6} 83 (1999).
\bibitem{Braghieri} A. Braghieri {\it et al.}, Phys. Lett. {\bf B363} 46 (1995).
\bibitem{Tejedor} J.A. Gomez Tejedor {\it et al.}, Nucl. Phys. {\bf A600} 413
(1996).
\bibitem{Haerter} F. Haerter {\it et al.}, Phys. Lett. {\bf B401} 229 (1997).
\bibitem{Zabrodin} A. Zabrodin {\it et al.}, Phys. Rev. {\bf C55} R1617 (1997);
                   Phys. Rev {\bf C60} 5201 (1999).
\bibitem{Krusche_4} B. Krusche {\it et al.}, Eur. Phys. J. {\bf A6} 309 (1999).
\bibitem{Kleber} V. Kleber {\it et al.}, Eur. Phys. J. {\bf A9} 1 (2000).
\bibitem{Wolf} M. Wolf {\it et al.}, Eur. Phys. J. {\bf A9} 5 (2000).
\bibitem{Tiator} L. Tiator {\it et al.}, Phys. Rev. {\bf C60} 35210 (1999).
\bibitem{Brown} G.E. Brown and M. Rho, Phys. Rev. Lett. {\bf 66} 2720 (1991).
\bibitem{Mosel}U. Mosel, Prog. Part. Nucl. Phys. 42, 163 (1999)
\bibitem{PDG} C. Caso {\it et al.}, Eur. Phys. J. {\bf C3} 1 (1998).
\bibitem{Frommhold} Th. Frommhold {\it et al.}, Phys. Lett. {\bf B295} 28
(1992); Z. Phys. {\bf A350} 249 (1994).
\bibitem{Bianchi} N. Bianchi {\it et al.}, Phys. Lett. {\bf B299} 219 (1993).
\bibitem{Lehr} J. Lehr {\it et al.}, Nucl. Phys. {\bf A671} 503
(2000).
\bibitem{Krusche_5} B. Krusche, Few-Body Suppl. 11 267 (1999).
\bibitem{Roebig} M. R\"obig-Landau {\it et al.}, Phys. Lett. {\bf B373} 45
(1996).
\bibitem{Effenberger} M. Effenberger {\it et al.}, Nucl. Phys. {\bf A614} (1997)
501.
\bibitem{Carrasco} R.C. Carrasco, Phys. Rev. {\bf C48} (1993) 2333.
\bibitem{Yorita} T. Yorita {\it et al.}, Phys. Lett. {\bf B476} 226 (2000)
\bibitem{MAID}
D. Drechsel {\it et al.}, Nucl. Phys. {\bf A645} 145 (1999),
http://www.kph.uni-mainz.de/MAID/maid.html. 
\bibitem{Lacombe} M. Lacombe {\it et al.}, Phys. Lett. {\bf B101} 139 (1981).
\bibitem{Hombach} A. Hombach {\it et al.}, Z. Phys. {\bf A 352} 223 (1995).
\bibitem{MacCormick} M. MacCormick {\it et al.}, Phys. Rev. {\bf C55} 1033
(1997).
\end{references}
\end{document}